\documentclass[superscriptaddress,showpacs,aps,twocolumn,floatfix,prl]{revtex4-1}
\usepackage{bm,amsmath,amssymb}
\usepackage{amssymb}
\usepackage{amsmath}
\usepackage{graphicx,bm}
\usepackage{verbatim}
\usepackage[dvipsnames]{xcolor}
\usepackage{soul}
\usepackage{ bbold }
\usepackage[colorlinks]{hyperref}
\AtBeginDocument{%
 \hypersetup{ 
 linkcolor=blue,
 filecolor=magenta, 
 urlcolor=blue,
 citecolor=blue,
 colorlinks=true,
 }
}

\usepackage[utf8]{inputenc} 
\DeclareUnicodeCharacter{0308}{\"{}} 

\begin{document}

\title{Quantum oscillation signatures of the Bloch-Gruneisen temperature in the Dirac semi-metal ZrTe$_5$}

\author{S. Galeski}
\affiliation{Physikalisches Institut, Universitat Bonn, Nussallee 12,53115 Bonn, Germany}
\affiliation{Hochfeld-Magnetlabor Dresden (HLD-EMFL) and Wurzburg-Dresden Cluster of Excellence ct.qmat, Helmholtz-Zentrum Dresden-Rossendorf, 01328, Dresden, Germany}

\author{K. Araki}
\affiliation{Applied Physics, National Defense Academy, Yokosuka, Kanagawa 239-8686, Japan}

\author{O. K. Forslund}
\affiliation{Physik-Institut, Universitat Zurich, Winterthurerstrasse 190, CH-8057 Zurich, Switzerland}
\affiliation{Department of Physics and Astronomy, Uppsala University, Box 516, SE-75120 Uppsala, Sweden}
 
 \author{R. Wawrzyńczak}
 \affiliation{Max Planck Institute for Chemical Physics of Solids, Nothnitzer Straße 40, 01187, Dresden, Germany}

\author{H.F. Legg}
\affiliation{Department of Physics, University of Basel, Klingelbergstrasse 82, CH-4056, Basel, Switzerland}

\author{P.K. Sivakumar}
\affiliation{Max Planck Institute of Microstructure Physics, Weinberg 2, 06120 Halle (Saale), Germany}

\author{U. Miniotaite}
\affiliation{Department of Applied Physics, KTH Royal Institute of Technology, SE-106 91 Stockholm, Sweden}

\author{F. Elson}
\affiliation{Department of Applied Physics, KTH Royal Institute of Technology, SE-106 91 Stockholm, Sweden}
 
\author{M. Månsson}
\affiliation{Department of Applied Physics, KTH Royal Institute of Technology, SE-106 91 Stockholm, Sweden}
 
\author{C. Witteveen}
\affiliation{Department of Quantum Matter Physics, University of Geneva, Quai Ernest-Ansermet 24, 1211 Geneva, Switzerland }
 
\author{F.O. von Rohr}
\affiliation{Department of Quantum Matter Physics, University of Geneva, Quai Ernest-Ansermet 24, 1211 Geneva, Switzerland }
 
\author{A.Q.R. Baron}
\affiliation{Precision Spectroscopy Division, CSRR, SPring-8/JASRI, Sayo, Hyogo 679-5198, Japan}
 
\author{D. Ishikawa}
\affiliation{Precision Spectroscopy Division, CSRR, SPring-8/JASRI, Sayo, Hyogo 679-5198, Japan}

\author{Q. Li}
\affiliation{Condensed Matter Physics and Materials Science Department, Brookhaven National Laboratory, Upton, NY, USA}
 
\author{G. Gu}
\affiliation{Condensed Matter Physics and Materials Science Department, Brookhaven National Laboratory, Upton, NY, USA}
 
\author{L. X. Zhao}
\affiliation{Beijing National Laboratory for Condensed Matter Physics, Institute of Physics, Chinese Academy of Sciences, 100190, Beijing, China}
\affiliation{Songshan Lake Materials Laboratory, 523808, Dongguan, Guangdong, China}
\affiliation{School of Physics Science, University of Chinese Academy of Sciences, 100049, Beijing, China}

\author{W. L. Zhu}
\affiliation{Beijing National Laboratory for Condensed Matter Physics, Institute of Physics, Chinese Academy of Sciences, 100190, Beijing, China}
\affiliation{School of Physics Science, University of Chinese Academy of Sciences, 100049, Beijing, China}
\affiliation{School of Physics and Information Technology, Shaanxi Normal University, 710062, Xi’an, China}

\author{G. F. Chen}
\affiliation{Beijing National Laboratory for Condensed Matter Physics, Institute of Physics, Chinese Academy of Sciences, 100190, Beijing, China}
\affiliation{Songshan Lake Materials Laboratory, 523808, Dongguan, Guangdong, China}
\affiliation{School of Physics Science, University of Chinese Academy of Sciences, 100049, Beijing, China}

\author{Y. Wang}
\affiliation{Beijing National Laboratory for Condensed Matter Physics, Institute of Physics, Chinese Academy of Sciences, 100190, Beijing, China}
 
 \author{S.S.P. Parkin}
 \affiliation{Max Planck Institute of Microstructure Physics, Weinberg 2, 06120 Halle (Saale), Germany}
 
 \author{D. Grobunov}
\affiliation{Hochfeld-Magnetlabor Dresden (HLD-EMFL) and Wurzburg-Dresden Cluster of Excellence ct.qmat, Helmholtz-Zentrum Dresden-Rossendorf, 01328, Dresden, Germany}
 
 \author{S. Zherlitsyn}
\affiliation{Hochfeld-Magnetlabor Dresden (HLD-EMFL) and Wurzburg-Dresden Cluster of Excellence ct.qmat, Helmholtz-Zentrum Dresden-Rossendorf, 01328, Dresden, Germany}
 
 \author{B. Vlaar}
\affiliation{Institute of Solid State Physics, TU Wien, Wiedner Hauptstr. 8-10, 1040 Vienna, Austria}
 
 \author{D. H. Nguyen}
\affiliation{Institute of Solid State Physics, TU Wien, Wiedner Hauptstr. 8-10, 1040 Vienna, Austria}
 
 \author{S. Paschen}
\affiliation{Institute of Solid State Physics, TU Wien, Wiedner Hauptstr. 8-10, 1040 Vienna, Austria}
 
 \author{P. Narang}
\affiliation{College of Letters and Science, University of California, Los Angeles, California 90095, USA}
 
 \author{C. Felser}
 \affiliation{Department of Physics, University of Basel, Klingelbergstrasse 82, CH-4056, Basel, Switzerland}
 
 \author{J. Wosnitza}
 \affiliation{Institut fur Festkorper- und Materialphysik, Technische Universitat Dresden, 01069, Dresden, Germany}
 \affiliation{Institute of Theoretical Physics and Wurzburg-Dresden Cluster of Excellence ct.qmat, Technische Universitat Dresden, 01069, Dresden, Germany}
 
 \author{T. Meng}
 \affiliation{Institute of Theoretical Physics and Wurzburg-Dresden Cluster of Excellence ct.qmat, Technische Universita ̈t Dresden, 01062 Dresden, Germany}

 \author{Y. Sassa}
 \affiliation{Department of Physics, Chalmers University of Technology, Kemigården 1, 412 96 Gothenburg, Sweden}

 \author{S.A. Hartnoll}
 \affiliation{Department of Applied Mathematics and Theoretical Physics, University of Cambridge, Cambridge, CB3 0WA, U.K.}

 \author{J. Gooth}
 \affiliation{Physikalisches Institut, Universitat Bonn, Nussallee 12,53115 Bonn, Germany}
 \affiliation{Max Planck Institute for Chemical Physics of Solids, Nothnitzer Straße 40, 01187, Dresden, Germany}

\date{\today}

\begin{abstract}
The electron-phonon interaction is in many ways a solid state equivalent of quantum electrodynamics. Being always present the e-p coupling is responsible for the intrinsic resistance of metals at finite temperatures making it one of the most fundamental interactions present in solids. In typical metals different regimes of e-p scattering are separated by a characteristic phonon energy scale – the Debye temperature. However in metals harboring very small Fermi surfaces, such as semimetals, a new scale emerges – the Bloch-Gruneisen temperature. A temperature at which the average phonon momentum becomes comparable to the Fermi momentum of the electrons. Here we report sub-kelvin transport and sound propagation experiments on the Dirac semimetal ZrTe$_5$. The combination of the simple band structure with only a single small Fermi surface sheet and low speed of sound allowed us to directly observe the Bloch-Gruneisen temperature and its consequences on electronic transport of a 3D metal in the limit where the small size of the Fermi surface leads to effective restoration of transnational invariance of free space. Our results indicate that on entering this hydrodynamic transport regime the viscosity of the Dirac electronic liquid undergoes an anomalous increase beyond the theoretically predicted T$^5$ temperature dependence. Extension of our measurements to strong magnetic fields reveal that despite the dimensional reduction of the electronic band structure the electronic liquid retains characteristics of the zero-field hydrodynamic regime up to the quantum limit. This is vividly reflected by an anomalous suppression of the amplitude of quantum oscillations seen in the Shubnikov-de Haas effect. 

\end{abstract}

\maketitle


The ability to conduct electrical current is one of the most basic characteristics of metals. In simple materials, where resistance is dominated by scattering of charge carriers by lattice vibrations, the temperature dependence of resistance falls into two regimes\cite{Ziman}: At high temperatures above the Debye Temperature $\Theta_D$, dynamics of the lattice can be described as an assembly of independent classical oscillators, giving rise to electron scattering rate that is linear in temperature. In contrast, at low temperature intrinsic resistance emerges from scattering of charge carriers with bosonic quasi-particles that represent the collective excitations of the lattice – phonons. Here only acoustic phonons with momentum smaller than $k_ph<(k_B T)⁄(\hbar v_S )$, where vs is the speed of sound, are populated and can take part in scattering, leading to a much stronger temperature dependence of resistance $R(T)\approx T^5$. The situation, however, is different in  materials hosting small Fermi Surfaces (FS) such as semimetals \cite{BehniaT2}\cite{BehniaAntimony}. In electron-phonon (e-p) collisions, the maximum allowed phonon scattering momentum is limited to $2k_F$ – twice the Fermi wave vector. This leads to a new temperature scale - the Bloch-Gruneisen temperature $\Theta_{BG }=2\hbar v_S k_F⁄k_B$ below which phonon scattering can no longer access the entire Fermi surface phase space, marking the crossover between $T^5$ and T-linear resistance. Furthermore it has been predicted that cooling below $\Theta_{BG }$ can lead to appearance of a range of novel phenomena including increase of the electron liquid viscosity and increase of the effective electron mass due to e-p interactions below this temperature  \cite{Hartnoll}\cite{Allen}. 

In this work we describe sub-kelvin charge transport and ultrasound propagation measurements on the Dirac semimetal ZrTe$_5$.Our results indicate an anomalous increase of electron viscosity on crossing $\Theta_{BG}$. Furthermore magnetoresistance measurements reveal an anomalous suppression of the Shubnikov-de Haas effect at sub-kelvin temperatures. Additional analysis of magneto-acoustics quantum oscillations (QO) and magnetoresistance in samples containing single and multiple FS sheets indicates that the anomalous behavior of SdH amplitudes is likely an electronic effect related to the suppression of Umklapp process and decoupling of the electron liquid from the lattice in low density samples. 

The pentatelluride material family (ZrTe$_5$ and HfTe$_5$) has risen to prominence in recent years as an excellent solid-state realization of the 3D Dirac Hamiltonian and material candidates to harbor a topological insulating ground state\cite{Mutch}\cite{Peng}.  Recent studies showed that, due to the exceptional purity (electron mobilities $\mu_e \approx$ 200 0000 cm2/Vs) most properties, down to 2 Kelvin, can be well accounted for using a simple low energy description\cite{Galeski1}\cite{Galeski2} making both ZrTe$_5$ and HfTe$_5$ \cite{DiSalvo}\cite{Galeski3}\cite{Wang} one of the best understood Dirac systems. ZrTe$_5$ is a diamagnetic metal and shows no signatures of magnetic interactions. While it has been shown to undergo a Lifshitz transition in the range of T$_L$=70-130K \cite{Skelton}\cite{Zhang1}\cite{Zhang2} , inducing a change in charge-carrier type, at low temperatures it is believed to be a simple metal.

Despite this apparent simplicity, the Fermi surface of the pentatellurides sensitively depends on the location of the chemical potential. Samples with small charge carrier density are characterized by a single elliptical electron-like Fermi surface at the gamma point covering only a few percent of the Brillion zone (BZ). Such a small FS together with the low sound  velocity of acoustic phonons provide an ideal setting for studying the crossover between different phonon scattering regimes with the estimated $\Theta_{BG}$ of only a few Kelvin. In contrast, in samples with a slightly higher charge carrier density can force the chemical potential to cross an additional bands giving rise to additional Fermi surface sheets in the corners of the BZ \cite{Zhang2}\cite{Facio}. The additional scattering channels available in those samples  provide a reference where physics related to the low $\Theta_{BG}$ temperature should be suppressed.

Here, we studied ZrTe$_5$ samples with different charge carrier densities to sub-Kelvin temperatures and examined their electrical transport in quantizing magnetic fields. For clarity of presentation in the remainder of the manuscript we primarily display data measured on two samples (A and B), differing in carrier density,  with data on additional samples presented in the supplementary material. In all samples, the four-terminal longitudinal electrical resistivity $\rho_xx$ and the Hall resistivity $\rho_{xy}$ were measured with the electrical current applied along the crystalographic a-axis and the magnetic field B applied along the b-axis, low field Hall effect measurements were used to extract the sample charge carrier densities.  

Results of magnetoresistance measurements on samples A, and B are shown in Figures 1 a-b. Both  samples display pronounced Shubnikov de-Haas oscillations exhibiting, however, very different frequencies. Comparison of the oscillations frequencies (Fig.1 c-d), with charge carrier density extracted from Hall measurements and is in good agreement with established band structure of ZrTe$_5$: increasing electron filling leads to expansion of the central Dirac FS and appearance of an additional oscillation frequency, reflecting that the chemical potential increases past the bottom of and additional quadratic band located at the corner of the BZ (for details and band structure calculation see the Supplementary Material).

\begin{figure}[ht!]
    \centering
   \includegraphics[width=0.5\textwidth]{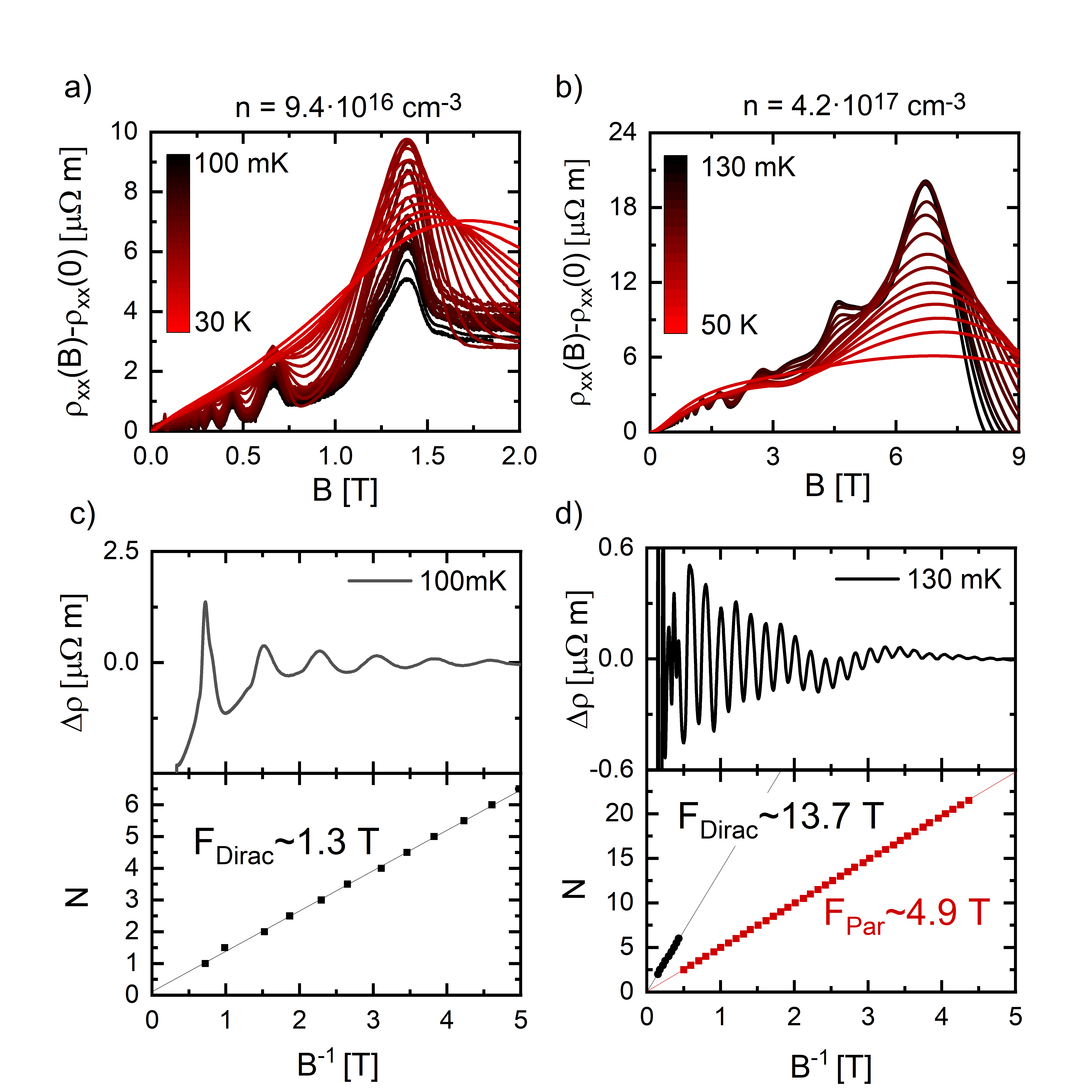}
    \caption{Shubnikov-de Haas oscillations in ZrTe5 in the sub-K range. a, Magnetic-field dependence of the longitudinal electrical resistivity $\rho_{xx}$ of sample A at various temperatures between 100 mK and 30 K with electrical current applied along the a-axis of the crystals and magnetic field applied along b. b, Magnetic-field dependence of the longitudinal electrical resistivity $\rho_{xx}$ of sample B at various temperatures between 130 mK and 50 K with electrical current applied along a and magnetic field applied along b. c, SdH oscillations of sample A plotted vs. inverse field together with the Landau fan diagram used for indexing the oscillations. d, SdH oscillations of sample B plotted vs. inverse field together with the Landau fan diagram used for indexing the oscillations originating from different FS pockets. }
    \label{fig1}
\end{figure}

Investigation of the temperature dependence of the  SdH oscillations amplitudes reveals an intriguing feature: for sample A the oscillation amplitude is strongly suppressed below about 2-3 K, Figure 2 a. In contrast in sample B, Figure 2b, the canonical Lifshitz- Kosevich(LK) behavior is recovered. In the context of the Bloch-Grunneisen temperature, suppression of SdH amplitude could be interpreted as originating from renormalization of the quasiparticle effective mass from the pure band mass m* to an effective mass renormalized by the e-p interaction: $m*=m_{Band}(1+\lambda)$ where $\lambda$ is the e-p coupling constant \cite{Shoenberg}\cite{Coleman}. Similar suppression of QO amplitude at low temperatures have been related to highly unconventional physics, including mass divergence close to quantum criticality \cite{McCollam}. Comparison of the Fermi velocities of sample A (sample B of ref \cite{Galeski2}) and the measured speed of longitudinal acoustic phonons (C11 mode) $v_a^T=3375$ m/s, allows to estimate the BG temperature: $\Theta_{BG} \approx 2.7$ K  - coinciding with the onset of the SdH oscillation amplitude suppression. However the magnitude of the SdH amplitude suppression would suggest an effective mass increase by more than a factor of two. This would require a very strong e-p coupling, in contrast with previous estimates\cite{Galeski2}\cite{Ehmcke}. 

\begin{figure}[ht!]
    \centering
   \includegraphics[width=0.52\textwidth]{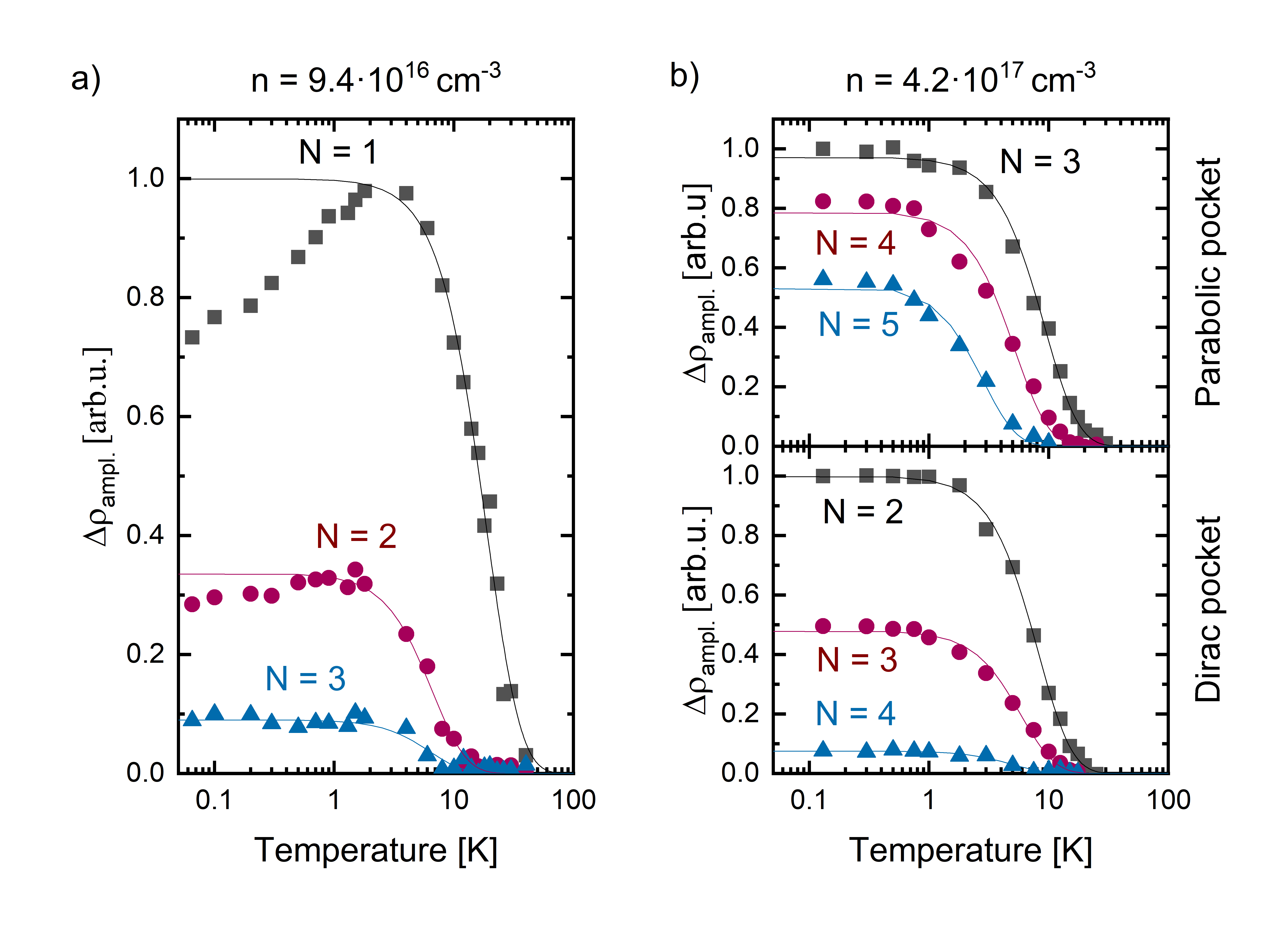}
    \caption{Figure 2  Temperature dependence of SdH oscillations of samples A and B. a, SdH amplitude as a function of T measured on sample A for three different magnetic fields. b, SdH amplitude for both FS pockets as a function of T measured on sample B for three magnetic fields. The dots are experimental data. The lines are fits using the Lifshitz-Kosevich (LK) formula taking into account data from above 2 K. }
    \label{fig2}
\end{figure}

In order to address this apparent contradiction we have resorted to a different probe of the quasi-particle effective mass. Although measurements of the SdH effect are the most widely used tool for the study of quantum oscillations in semimetals, their interpretation can in principle be much more difficult than other, thermodynamic, probes. The difficulty lies in the fact that magnetoresistance measurements on their own are a probe of momentum relaxation. Typically, in such measurements, one assumes that at low enough temperatures momentum scattering becomes temperature and field independent and thus the SdH oscillation amplitude  reflects the field induced oscillations of the density of states and thus the effective mass\cite{Shoenberg}. This, however, does not necessarily have to be the case. In order to verify whether the suppression of QO in ZrTe$_5$ can be interpreted as an enhancement of the effective mass we measured the temperature dependence of QO seen in the speed of sound. Velocity of sound  is directly related to the system’s elastic modulus and thus is a thermodynamic quantity that can be expected to directly reflect the field and temperature dependence of the free energy without spurious scattering effects\cite{LeBoeuf}\cite{Luthi}.

Ultrasonic measurements require special sample surface preparation thus we have selected an additional sample - C from the same growth batch as sample A in order to leave sample A intact for future measurements. Figure 3a shows the low temperature field dependence of the speed of sound of the C11 mode with magnetic field applied parallel to the b-axis. Analysis of the frequency of magnetoacoustic oscillations measured in sample C reveals it enters the quantum limit at a similar field as sample A, confirming that both samples share similar charge carrier density. However, in contrast to the SdH oscillations seen in sample A  (Figure 3b) the amplitude of the magnetoacoustic quantum oscillations in sample C does not reveal any amplitude suppression at low temperature. Instead the oscillation amplitude saturates following the standard LK behaviour, Figure 3c. The discrepancy in the behaviour of the QO observed in thermodynamic and transport measurements suggests that the anomalous SdH amplitude originates from a low temperature change in the momentum relaxation of the electron fluid.

\begin{figure}[ht!]
    \centering
   \includegraphics[width=0.52\textwidth]{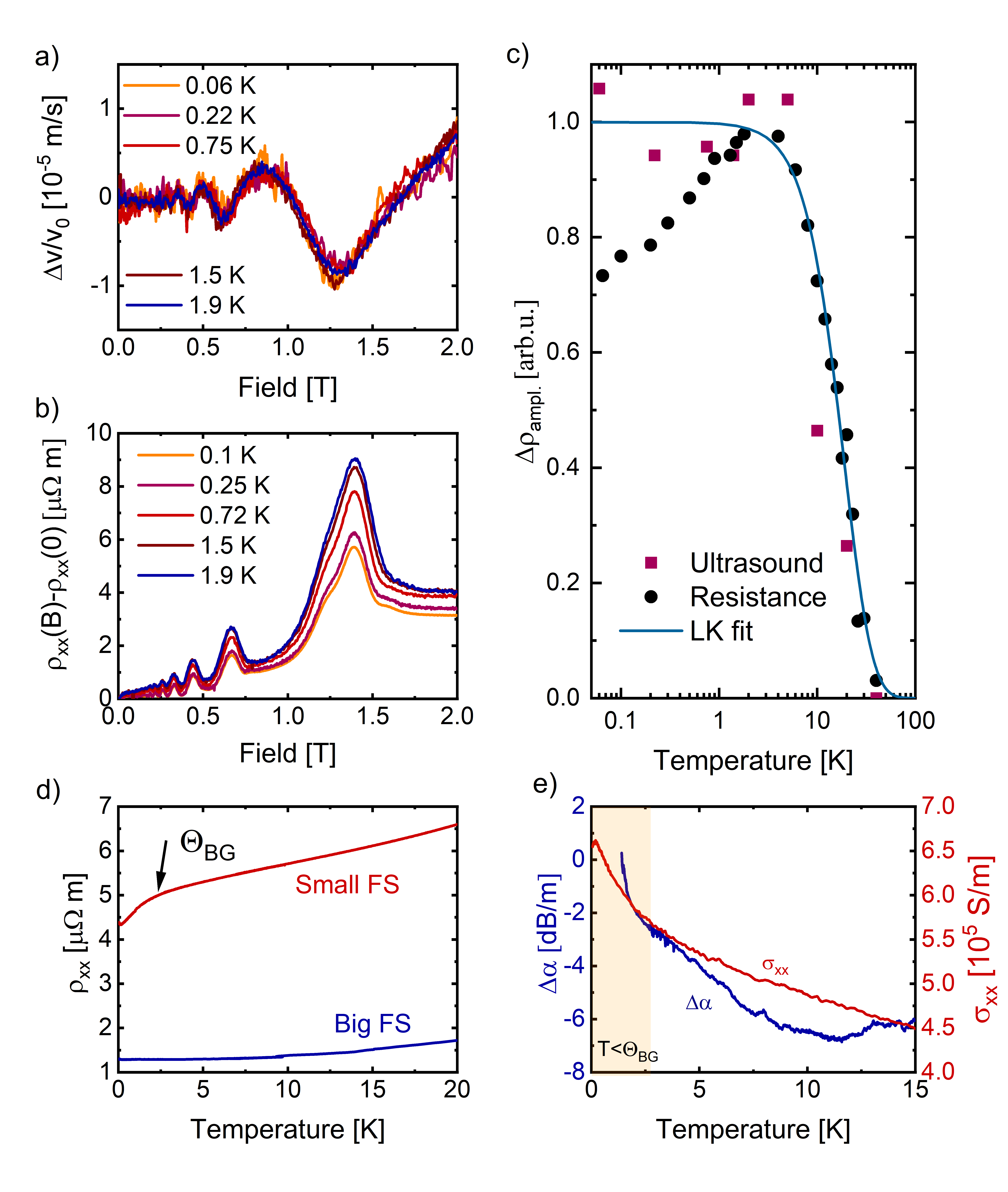}
    \caption{Comparison of resistance measurements with sound propagation and attenuation measurements.  a, Magneto-acoustic quantum oscillations measured in sample C with a longitudinal sound mode propagating along the a-axis and field applied along the b-axis. b, SdH effect measured in sample A (see also Figure 1a). c, Comparison of QO amplitudes obtained from speed-of-sound and magnetoresistance measurements. d, Comparison of zero-field resistances measured in sample A and B. Sample A with a small FS exhibits an anomalous downturn at low temperatures. The resistance in sample B, in contrast, saturates at low T. e, Comparison of the zero-field electrical conductivity measured in sample A with the sound attenuation measured in sample C. Both quantities exhibit an anomalous increase on crossing $\Theta_{BG}$  with a stronger effect seen in the sound attenuation reflecting a rapid increase of the electron-liquid viscosity. }
    \label{fig3}
\end{figure}

\twocolumngrid\

In the absence of a lattice, all electron-electron (e-e) collisions conserve momentum. In such circumstance a single component electron fluid cannot dissipate its momentum. However, the presence of the lattice changes the type of possible scattering events introducing so-called Umklapp processes, where collisions scatter one of the electrons to a neighbouring BZ allowing to transfer a reciprocal lattice vector of momentum (G) to the lattice. Such processes, although very efficient in relaxing momentum, are only possible if the FS spans at least half of the BZ, and are strongly suppressed for smaller FSs \cite{BehniaT2}3. Thus, for small FSs such as those found in samples A and C where the FS comprises less than 5\% of the BZ \cite{Galeski2}, the main route for momentum relaxation, in the absence of disorder, leads through electron-phonon (e-p) collisions where the electron liquid leaks momentum to the phonon gas. Consequently, the suppression of the SdH oscillations seems to emerge from suppression of e-p collisions below the BG temperature. This interpretation is corroborated by  the absence of the SdH amplitude suppression in the sample B consisting of several FS sheets where Umklapp processes are expected to play an important role in momentum dissipation. 

Additional evidence for the influence of the BG temperature on the low temperature properties of ZrTe$_5$ stems from the zero-field temperature dependence of resistivity. Figure 3d shows the temperature dependence of resistivity of samples A and B. In sample B, with high charge carrier density, the low temperature resistance tends to saturate following the canonical behaviour expected from the low temperature resistance of metals $R(T)=R0+AT^2+BT^5$ \cite{Ziman} with the quadratic term originating from e-e scattering events, the T$5$ term from e-p scattering and R0 being the residual resistivity due to impurities and imperfections1. In contrast, the resistance of sample A hosting only a single small FS undergoes a highly anomalous transition with a rapid downturn on crossing $\Theta_{BG}$. Before commenting further on this downturn, we describe an analogous anomaly in the sound attenuation.

 The temperature dependence of sound attenuation is shown in sample C, see figure 3e. Unlike sound velocity which is a thermodynamic probe, sound attenuation is a probe of energy transfer akin to thermal conductivity \cite{Holland}. In our ultrasound experiments we used a 314 MHz excitation, this translates to a wavelength of ca 10 $\mu$ m, an order of magnitude higher than the electron mean free path estimated from the Drude formula at 2K, $l_{mfp} \approx 1.5\mu$ m. In this limit the ultrasonic attenuation is proportional to the electronic liquid viscosity $\nu$.\cite{Allen}\cite{Luthi}. It can be seen in figure 3e that the measured attenuation exhibits a sudden increase on crossing $\Theta_{BG}$. Interestingly the increase of electron viscosity is much sharper than that seen in conductivity. An increase of electron viscosity is expected once e-p scattering is suppressed, however Bloch-Grunneisen theory suggests that both conductivity and viscosity should follow a T$^{-1}$ dependence at high temperatures and T$^5$ below $\Theta_{BG}$\cite{Allen}. Although in the present case at high temperatures the dependence is roughly linear it decreases much faster than the predicted T$^{-5}$ law at low temperatures. Notably, the resistivity has the opposite curvature around $\Theta_{BG}$  to what is conventionally expected. Here we would like to highlight the robustness of the effect: the same effect is observed in two different Te-flux grown samples, using measurement methods based on distinctively different physical phenomena (conductivity, sound attenuation). In addition we have found a similar effect in CVT grown HfTe$_5$ samples with charge carrier densities similar to samples A and C (for details see SI). Our experiments thus demonstrate that the observed effect is intrinsic and related to a low $\Theta_{BG}$, small charge carrier density and lack of Umklapp scattering processes. One obvious suspect in such conditions could be phonon-drag. However comparison of the measured data with a variational calculation of e-p scattering rates (see SI for details) that take into account details of the Fermi surface geometry, phonon velocity anisotropies and phonon drag excludes such an origin.

Recently the low temperature resistivity of small Fermi surface metals has been discussed in the context of  Bi$_2$O$_2$Se and SrTiO$_3$ 3,21\cite{BehniaT2}\cite{Wang2}. Here it was shown that resistance follows the canonical Fermi liquid predictions with  $R(T) \approx T^2$ at low temperatures. Although the $T^2$ resistivity law is usually taken as a hallmark of Fermi liquid behavior it is difficult to detect in common metals where resistance is dominated by the T$_5$ term originating from e-p processes\cite{Behnia}. In the case of low density metals, the increased e-e scattering cross section is expected to enhance the contribution of the $T^2$ term. However for the $T^2$ term to be observable there should be an additional process allowing the electron liquid to leak momentum (such as multiple FS with mismatched effective mases, Umklapp processes)\cite{Wang}\cite{Behnia}. It has been argued that the electrical resistance of SrTiO$_3$ is instead due to two-phonon scattering\cite{Kumar}. 

In our measurements we have not detected the T$^2$ resistance to the lowest measured temperatures in low density samples with a single component FS. In our measurements we have focused on samples with a charge carrier density four times smaller than in previous experiments on Bi2O2Se and SrTiO$_3$\cite{BehniaT2}\cite{Wang2}. This combined with the small phonon velocity conspires to place our low density samples in a very different regime, to our knowledge, one with the lowest observed $\Theta_{BG}$  allowing to observe transport across the $\Theta_{BG}$  without the contribution of Umklapp processes and in the presence of Landau levels. 

Our experiments suggest a rapid increase of the electron fluid viscosity that could manifest an onset of a regime of viscous hydrodynamics. This could be qualitatively understood by comparing the frequency of momentum relaxing scattering with the quantum coherence time obtained from the field dependance of quantum oscillations\cite{Moll}. The momentum relaxation scattering time  can be estimated from the zero-field resistance using the Drude formula. At $\approx$2K using the parameters of sample A one obtains $\tau_e$=5.4ps. In contrast the quantum coherence time previously extracted from the SdH effect on sample A (see sample B of ref \cite{Galeski2}) at the same temperature, depending on the field orientation, is in the range 0.2-0.9 ps. This suggests that the dynamics of the electrons in ZrTe$_5$ at low temperatures has a strong contribution from normal e-e collisions with rare momentum relaxing collisions. 

We have already seen that crossing $\Theta_{BG}$  and entering the regime of increased viscosity has a profound impact on the SdH oscillations. Indeed comparison of the zero field resistance (figure 3d) with the temperature dependance of the quantum oscillations in figure 1a reveals that the resistance at the maxima of the SdH oscillations is suppressed more strongly than zero field resistance. This observation, together with the fact that the temperature dependance of quantum oscillation represents the Fourier transform of the electron distribution function in energy\cite{Sebastian}, could suggest that on cooling through $\Theta_{BG}$  electrons decouple from the thermal bath and are able remain in a non-equilibrium distribution. Such a scenario would be in qualitative agreement with SdH amplitudes deviating from the LK formula due to a continuous pumping of momentum to the electron liquid by the electric field and with thermodynamic magneto acoustic oscillations exhibiting standard behavior. However, confirmation of this exciting possibility requires further theoretical work and experiments such as a direct measurement of the temperature difference between the electron liquid and the lattice as temperature changes across $\Theta_{BG}$. 

 To date electron hydrodynamics has been studied in several 3D compounds notably WTe$_2$\cite{Vool}, WP$_2$\cite{Jogi}\cite{Pri}, PdCoO$_2$\cite{Moll} and elemental antimony\cite{BehniaAntimony}. In these studies signatures of hydrodynamic flow have been primarily inferred from the influence of boundary scattering on transport properties. However to date the temperature dependence of viscosity of the electron liquid on entering the hydrodynamic regime has not been directly probed. In the light of the anomalous response seen in both conductivity and sound attenuation in our experiments we believe it is crucial to investigate sound attenuation in those compounds across different predicted regimes29\cite{Lucas}. Here an interesting issue is that in most theoretical treatments of electron hydrodynamics and  e-p coupling it is assumed that the speed of sound is much smaller than the Fermi velocity and interactions are short range. This is contrary to the conditions found in ZrTe$_5$ since the ratio of the speed of sound and along the b-axis obtained from inelastic X-ray scattering on ZrTe$_5$ (for details see Supplementary Material) and the Fermi velocity is of the order of ~0.2 and thus beyond usual assumptions. In additional our estimate of the Thomas-Fermi screening length indicates that the screening length exceeds several unit cells and thus cannot be treated as short ranged (for details see Supplementary Material). 

 In summary in we have identified a possible onset of an electron hydrodynamic regime in the Dirac semimetal ZrTe$_5$. Our measurements point to an anomalous enhancement of electron viscosity and conductivity below the  $\Theta_{BG}$  temperature. Furthermore we have demonstrated that the relevant temperature scales of the pentatellurides enable the study of a novel and thus far inaccessible regime where the presence of quantum oscillations overlaps with the hydrodynamic regime. To our knowledge this regime has not been studied neither experimentally nor theoretically. In particular the possibility of easy exfoliation and production of field-gatable devices with sub-micron size makes ZrTe$_5$ an excellent testing platform allowing for excellent level of control of channel size and carrier concentration to study effects of electron viscosity\cite{Liu}.

\nocite{*}

\begin{acknowledgments}
H.F.L. acknowledges funding by the Georg H. Endress foundation. O.K.F acknowledges funding by the Swedish Research Council (VR) via a Grant 2022-06217 and the Foundation Blanceflor 2023 fellow scholarship. TM acknowledges funding by the Deutsche Forschungsgemeinschaft (DFG) via the Emmy Noether Programme (Quantum Design grant, ME4844/1, project- id 327807255), project A04 of the Collaborative Research Center SFB 1143 (project-id 247310070), the Cluster of Excellence on Complexity and Topology in Quantum Matter ct.qmat (EXC 2147, project-id 390858490), as well as from the Luxembourg National Research Fund (FNR) and the DFG through the CORE grant “Topology in relativistic semimetals (TOPREL)” (FNR project No. C20/MS/14764976 and DFG Project No. 452557895). Y.W. acknowledges funding support from the Chinese Academy of Sciences (Nos. YSBR047 and E2K5071). We acknowledge the support of the Hochfeld-Magnetlabor Dresden (HLD) at HZDR, member of the European Magnetic Field Laboratory (EMFL), the DFG through the Collaborative Research Center SFB 1143 (Project No. 247310070), and the Wurzburg-Dresden Cluster of Excellence ct.qmat (EXC 2147, Project No. 39085490). S.G., J.G., B.V., D.H.N., and S.P. acknowledge funding by the European Union's Horizon 2020 Research and Innovation Programme (Grant Agreement no 824109, the European Microkelvin Platform), S.G., J.G., and C.F. by the DFG through the Forschergruppe FOR 5249 (QUAST), and D.H.N. and S.P. by the Austria Science Fund (FWF) through I 5868-N (FOR 5249, QUAST).

\end{acknowledgments}

\bibliography{apssamp.bib}

\begin{thebibliography}{33}%
\makeatletter
\providecommand \@ifxundefined [1]{%
 \@ifx{#1\undefined}
}%
\providecommand \@ifnum [1]{%
 \ifnum #1\expandafter \@firstoftwo
 \else \expandafter \@secondoftwo
 \fi
}%
\providecommand \@ifx [1]{%
 \ifx #1\expandafter \@firstoftwo
 \else \expandafter \@secondoftwo
 \fi
}%
\providecommand \natexlab [1]{#1}%
\providecommand \enquote  [1]{``#1''}%
\providecommand \bibnamefont  [1]{#1}%
\providecommand \bibfnamefont [1]{#1}%
\providecommand \citenamefont [1]{#1}%
\providecommand \href@noop [0]{\@secondoftwo}%
\providecommand \href [0]{\begingroup \@sanitize@url \@href}%
\providecommand \@href[1]{\@@startlink{#1}\@@href}%
\providecommand \@@href[1]{\endgroup#1\@@endlink}%
\providecommand \@sanitize@url [0]{\catcode `\\12\catcode `\$12\catcode `\&12\catcode `\#12\catcode `\^12\catcode `\_12\catcode `\%12\relax}%
\providecommand \@@startlink[1]{}%
\providecommand \@@endlink[0]{}%
\providecommand \url  [0]{\begingroup\@sanitize@url \@url }%
\providecommand \@url [1]{\endgroup\@href {#1}{\urlprefix }}%
\providecommand \urlprefix  [0]{URL }%
\providecommand \Eprint [0]{\href }%
\providecommand \doibase [0]{https://doi.org/}%
\providecommand \selectlanguage [0]{\@gobble}%
\providecommand \bibinfo  [0]{\@secondoftwo}%
\providecommand \bibfield  [0]{\@secondoftwo}%
\providecommand \translation [1]{[#1]}%
\providecommand \BibitemOpen [0]{}%
\providecommand \bibitemStop [0]{}%
\providecommand \bibitemNoStop [0]{.\EOS\space}%
\providecommand \EOS [0]{\spacefactor3000\relax}%
\providecommand \BibitemShut  [1]{\csname bibitem#1\endcsname}%
\let\auto@bib@innerbib\@empty
\bibitem [{\citenamefont {Ziman}(2001)}]{Ziman}%
  \BibitemOpen
  \bibfield  {author} {\bibinfo {author} {\bibfnamefont {J.~M.}\ \bibnamefont {Ziman}},\ }\href@noop {} {\emph {\bibinfo {title} {Electrons and Phonons: The Theory of Transport Phenomena in Solids}}}\ (\bibinfo  {publisher} {Oxford University Press},\ \bibinfo {year} {2001})\BibitemShut {NoStop}%
\bibitem [{\citenamefont {Lin}\ \emph {et~al.}(2015)\citenamefont {Lin}, \citenamefont {Fauqué},\ and\ \citenamefont {Behnia}}]{BehniaT2}%
  \BibitemOpen
  \bibfield  {author} {\bibinfo {author} {\bibfnamefont {X.}~\bibnamefont {Lin}}, \bibinfo {author} {\bibfnamefont {B.}~\bibnamefont {Fauqué}},\ and\ \bibinfo {author} {\bibfnamefont {K.}~\bibnamefont {Behnia}},\ }\bibfield  {title} {\bibinfo {title} {Scalable $t^2$ resistivity in a small single-component fermi surface},\ }\href@noop {} {\bibfield  {journal} {\bibinfo  {journal} {Science}\ }\textbf {\bibinfo {volume} {349}},\ \bibinfo {pages} {945–948} (\bibinfo {year} {2015})}\BibitemShut {NoStop}%
\bibitem [{\citenamefont {Jaoui}\ \emph {et~al.}(2021)\citenamefont {Jaoui}, \citenamefont {Fauqué},\ and\ \citenamefont {Behnia}}]{BehniaAntimony}%
  \BibitemOpen
  \bibfield  {author} {\bibinfo {author} {\bibfnamefont {A.}~\bibnamefont {Jaoui}}, \bibinfo {author} {\bibfnamefont {B.}~\bibnamefont {Fauqué}},\ and\ \bibinfo {author} {\bibfnamefont {K.}~\bibnamefont {Behnia}},\ }\bibfield  {title} {\bibinfo {title} {Thermal resistivity and hydrodynamics of the degenerate electron fluid in antimony.},\ }\href@noop {} {\bibfield  {journal} {\bibinfo  {journal} {Nat. Commun.}\ }\textbf {\bibinfo {volume} {12}},\ \bibinfo {pages} {195} (\bibinfo {year} {2021})}\BibitemShut {NoStop}%
\bibitem [{\citenamefont {Hartnoll}\ and\ \citenamefont {Mackenzie}(2022)}]{Hartnoll}%
  \BibitemOpen
  \bibfield  {author} {\bibinfo {author} {\bibfnamefont {S.~A.}\ \bibnamefont {Hartnoll}}\ and\ \bibinfo {author} {\bibfnamefont {A.}~\bibnamefont {Mackenzie}},\ }\bibfield  {title} {\bibinfo {title} {Colloquium: Planckian dissipation in metals},\ }\href@noop {} {\bibfield  {journal} {\bibinfo  {journal} {Rev. Mod. Phys.}\ }\textbf {\bibinfo {volume} {11}},\ \bibinfo {pages} {94} (\bibinfo {year} {2022})}\BibitemShut {NoStop}%
\bibitem [{\citenamefont {Khan}\ and\ \citenamefont {Allen}(1987)}]{Allen}%
  \BibitemOpen
  \bibfield  {author} {\bibinfo {author} {\bibfnamefont {F.~S.}\ \bibnamefont {Khan}}\ and\ \bibinfo {author} {\bibfnamefont {P.~B.}\ \bibnamefont {Allen}},\ }\bibfield  {title} {\bibinfo {title} {Sound attenuation by electrons in metals},\ }\href@noop {} {\bibfield  {journal} {\bibinfo  {journal} {Phys. Rev. B}\ }\textbf {\bibinfo {volume} {35}},\ \bibinfo {pages} {1002} (\bibinfo {year} {1987})}\BibitemShut {NoStop}%
\bibitem [{\citenamefont {Mutch}\ \emph {et~al.}(2019)\citenamefont {Mutch} \emph {et~al.}}]{Mutch}%
  \BibitemOpen
  \bibfield  {author} {\bibinfo {author} {\bibfnamefont {J.}~\bibnamefont {Mutch}} \emph {et~al.},\ }\bibfield  {title} {\bibinfo {title} {Evidence for a strain-tuned topological phase transition in zrte$_5$},\ }\href@noop {} {\bibfield  {journal} {\bibinfo  {journal} {Sci. Adv.}\ }\textbf {\bibinfo {volume} {5}},\ \bibinfo {pages} {eaav9771} (\bibinfo {year} {2019})}\BibitemShut {NoStop}%
\bibitem [{\citenamefont {Zhang}\ \emph {et~al.}(2021)\citenamefont {Zhang} \emph {et~al.}}]{Peng}%
  \BibitemOpen
  \bibfield  {author} {\bibinfo {author} {\bibfnamefont {P.}~\bibnamefont {Zhang}} \emph {et~al.},\ }\bibfield  {title} {\bibinfo {title} {Observation and control of the weak topological insulator state in zrte$_5$},\ }\href@noop {} {\bibfield  {journal} {\bibinfo  {journal} {Nat. Commun.}\ }\textbf {\bibinfo {volume} {12}},\ \bibinfo {pages} {406} (\bibinfo {year} {2021})}\BibitemShut {NoStop}%
\bibitem [{\citenamefont {Galeski}\ \emph {et~al.}(2022)\citenamefont {Galeski} \emph {et~al.}}]{Galeski1}%
  \BibitemOpen
  \bibfield  {author} {\bibinfo {author} {\bibfnamefont {S.}~\bibnamefont {Galeski}} \emph {et~al.},\ }\bibfield  {title} {\bibinfo {title} {Signatures of a magnetic-field-induced lifshitz transition in the ultra-quantum limit of the topological semimetal zrte$_5$},\ }\href@noop {} {\bibfield  {journal} {\bibinfo  {journal} {Nat. Commun.}\ }\textbf {\bibinfo {volume} {13}},\ \bibinfo {pages} {7418} (\bibinfo {year} {2022})}\BibitemShut {NoStop}%
\bibitem [{\citenamefont {Galeski}\ \emph {et~al.}(2020)\citenamefont {Galeski} \emph {et~al.}}]{Galeski2}%
  \BibitemOpen
  \bibfield  {author} {\bibinfo {author} {\bibfnamefont {S.}~\bibnamefont {Galeski}} \emph {et~al.},\ }\bibfield  {title} {\bibinfo {title} {Origin of the quasi-quantized hall effect in zrte$_5$},\ }\href@noop {} {\bibfield  {journal} {\bibinfo  {journal} {Nat. Commun.}\ }\textbf {\bibinfo {volume} {12}},\ \bibinfo {pages} {3197} (\bibinfo {year} {2020})}\BibitemShut {NoStop}%
\bibitem [{\citenamefont {DiSalvo}\ \emph {et~al.}(1981)\citenamefont {DiSalvo}, \citenamefont {Fleming},\ and\ \citenamefont {Waszczak}}]{DiSalvo}%
  \BibitemOpen
  \bibfield  {author} {\bibinfo {author} {\bibfnamefont {F.~J.}\ \bibnamefont {DiSalvo}}, \bibinfo {author} {\bibfnamefont {R.~M.}\ \bibnamefont {Fleming}},\ and\ \bibinfo {author} {\bibfnamefont {J.~V.}\ \bibnamefont {Waszczak}},\ }\bibfield  {title} {\bibinfo {title} {Possible phase transition in the quasi-one-dimensional materials zrte$_5$ or hfte$_5$.},\ }\href@noop {} {\bibfield  {journal} {\bibinfo  {journal} {Phys. Rev. B}\ }\textbf {\bibinfo {volume} {24}},\ \bibinfo {pages} {2935} (\bibinfo {year} {1981})}\BibitemShut {NoStop}%
\bibitem [{\citenamefont {Galeski}\ \emph {et~al.}(2021)\citenamefont {Galeski} \emph {et~al.}}]{Galeski3}%
  \BibitemOpen
  \bibfield  {author} {\bibinfo {author} {\bibfnamefont {S.}~\bibnamefont {Galeski}} \emph {et~al.},\ }\bibfield  {title} {\bibinfo {title} {Unconventional hall response in the quantum limit of hfte$_5$},\ }\href@noop {} {\bibfield  {journal} {\bibinfo  {journal} {Nat. Commun.}\ }\textbf {\bibinfo {volume} {11}},\ \bibinfo {pages} {5926} (\bibinfo {year} {2021})}\BibitemShut {NoStop}%
\bibitem [{\citenamefont {Wang}(2021)}]{Wang}%
  \BibitemOpen
  \bibfield  {author} {\bibinfo {author} {\bibfnamefont {C.}~\bibnamefont {Wang}},\ }\bibfield  {title} {\bibinfo {title} {Thermodynamically induced transport anomaly in dilute metals zrte$_5$ and hfte$_5$},\ }\href@noop {} {\bibfield  {journal} {\bibinfo  {journal} {Phys. Rev. Lett.}\ }\textbf {\bibinfo {volume} {126}},\ \bibinfo {pages} {126601} (\bibinfo {year} {2021})}\BibitemShut {NoStop}%
\bibitem [{\citenamefont {Skelton}\ \emph {et~al.}(1981)\citenamefont {Skelton} \emph {et~al.}}]{Skelton}%
  \BibitemOpen
  \bibfield  {author} {\bibinfo {author} {\bibfnamefont {E.~F.}\ \bibnamefont {Skelton}} \emph {et~al.},\ }\bibfield  {title} {\bibinfo {title} {Giant resistivity and x-ray diffraction anomalies in low-dimensional zrte$_5$ and hfte$_5$},\ }\href@noop {} {\bibfield  {journal} {\bibinfo  {journal} {Solid State Commun}\ }\textbf {\bibinfo {volume} {42}},\ \bibinfo {pages} {1} (\bibinfo {year} {1981})}\BibitemShut {NoStop}%
\bibitem [{\citenamefont {Zhang}\ \emph {et~al.}(2017{\natexlab{a}})\citenamefont {Zhang} \emph {et~al.}}]{Zhang1}%
  \BibitemOpen
  \bibfield  {author} {\bibinfo {author} {\bibfnamefont {Y.}~\bibnamefont {Zhang}} \emph {et~al.},\ }\bibfield  {title} {\bibinfo {title} {Temperature-induced lifshitz transition in topological insulator candidate hfte$_5$},\ }\href@noop {} {\bibfield  {journal} {\bibinfo  {journal} {Sci. Bull.}\ }\textbf {\bibinfo {volume} {62}},\ \bibinfo {pages} {950–956} (\bibinfo {year} {2017}{\natexlab{a}})}\BibitemShut {NoStop}%
\bibitem [{\citenamefont {Zhang}\ \emph {et~al.}(2017{\natexlab{b}})\citenamefont {Zhang} \emph {et~al.}}]{Zhang2}%
  \BibitemOpen
  \bibfield  {author} {\bibinfo {author} {\bibfnamefont {Y.}~\bibnamefont {Zhang}} \emph {et~al.},\ }\bibfield  {title} {\bibinfo {title} {Electronic evidence of temperature-induced lifshitz transition and topological nature in zrte$_5$},\ }\href@noop {} {\bibfield  {journal} {\bibinfo  {journal} {Nat. Commun.}\ }\textbf {\bibinfo {volume} {8}},\ \bibinfo {pages} {15512} (\bibinfo {year} {2017}{\natexlab{b}})}\BibitemShut {NoStop}%
\bibitem [{\citenamefont {Facio}\ \emph {et~al.}(2023)\citenamefont {Facio} \emph {et~al.}}]{Facio}%
  \BibitemOpen
  \bibfield  {author} {\bibinfo {author} {\bibfnamefont {J.~I.}\ \bibnamefont {Facio}} \emph {et~al.},\ }\bibfield  {title} {\bibinfo {title} {Engineering a pure dirac regime in zrte$_5$},\ }\href@noop {} {\bibfield  {journal} {\bibinfo  {journal} {SciPost Phys.}\ }\textbf {\bibinfo {volume} {14}},\ \bibinfo {pages} {66} (\bibinfo {year} {2023})}\BibitemShut {NoStop}%
\bibitem [{\citenamefont {Shoenberg}(1984)}]{Shoenberg}%
  \BibitemOpen
  \bibfield  {author} {\bibinfo {author} {\bibfnamefont {D.}~\bibnamefont {Shoenberg}},\ }\href@noop {} {\emph {\bibinfo {title} {Magnetic Oscillations in Metals}}}\ (\bibinfo  {publisher} {Cambridge University Press},\ \bibinfo {year} {1984})\BibitemShut {NoStop}%
\bibitem [{\citenamefont {Coleman}(2015)}]{Coleman}%
  \BibitemOpen
  \bibfield  {author} {\bibinfo {author} {\bibfnamefont {P.}~\bibnamefont {Coleman}},\ }\href@noop {} {\emph {\bibinfo {title} {Introduction to Many-Body Physics}}}\ (\bibinfo  {publisher} {Cambridge University Press},\ \bibinfo {year} {2015})\BibitemShut {NoStop}%
\bibitem [{\citenamefont {McCollam}\ \emph {et~al.}(2005)\citenamefont {McCollam} \emph {et~al.}}]{McCollam}%
  \BibitemOpen
  \bibfield  {author} {\bibinfo {author} {\bibfnamefont {A.}~\bibnamefont {McCollam}} \emph {et~al.},\ }\bibfield  {title} {\bibinfo {title} {Anomalous de haas-van alphen oscillations in cecoin5$_5$},\ }\href@noop {} {\bibfield  {journal} {\bibinfo  {journal} {Phys. Rev. Lett.}\ }\textbf {\bibinfo {volume} {94}},\ \bibinfo {pages} {186401} (\bibinfo {year} {2005})}\BibitemShut {NoStop}%
\bibitem [{\citenamefont {Ehmcke}\ \emph {et~al.}(2021)\citenamefont {Ehmcke} \emph {et~al.}}]{Ehmcke}%
  \BibitemOpen
  \bibfield  {author} {\bibinfo {author} {\bibfnamefont {T.}~\bibnamefont {Ehmcke}} \emph {et~al.},\ }\bibfield  {title} {\bibinfo {title} {Propagation of longitudinal acoustic phonons in zrte5 exposed to a quantizing magnetic field},\ }\href@noop {} {\bibfield  {journal} {\bibinfo  {journal} {Phys. Rev. B}\ }\textbf {\bibinfo {volume} {104}},\ \bibinfo {pages} {245117} (\bibinfo {year} {2021})}\BibitemShut {NoStop}%
\bibitem [{\citenamefont {LeBoeuf}\ \emph {et~al.}(2017)\citenamefont {LeBoeuf} \emph {et~al.}}]{LeBoeuf}%
  \BibitemOpen
  \bibfield  {author} {\bibinfo {author} {\bibfnamefont {D.}~\bibnamefont {LeBoeuf}} \emph {et~al.},\ }\bibfield  {title} {\bibinfo {title} {Thermodynamic signatures of the field-induced states of graphite},\ }\href@noop {} {\bibfield  {journal} {\bibinfo  {journal} {Nat. Commun.}\ }\textbf {\bibinfo {volume} {8}},\ \bibinfo {pages} {1337} (\bibinfo {year} {2017})}\BibitemShut {NoStop}%
\bibitem [{\citenamefont {Luthi}(2005)}]{Luthi}%
  \BibitemOpen
  \bibfield  {author} {\bibinfo {author} {\bibfnamefont {B.}~\bibnamefont {Luthi}},\ }\href@noop {} {\emph {\bibinfo {title} {Physical Acoustics in the Solid State}}}\ (\bibinfo  {publisher} {Springer},\ \bibinfo {year} {2005})\BibitemShut {NoStop}%
\bibitem [{\citenamefont {Holland}(1968)}]{Holland}%
  \BibitemOpen
  \bibfield  {author} {\bibinfo {author} {\bibfnamefont {M.}~\bibnamefont {Holland}},\ }\bibfield  {title} {\bibinfo {title} {Thermal conductivity and ultrasonic attenuation},\ }\href@noop {} {\bibfield  {journal} {\bibinfo  {journal} {IEEE Trans. on Sonics Ultrason.}\ }\textbf {\bibinfo {volume} {15}},\ \bibinfo {pages} {18} (\bibinfo {year} {1968})}\BibitemShut {NoStop}%
\bibitem [{\citenamefont {Wang}\ \emph {et~al.}(2020)\citenamefont {Wang} \emph {et~al.}}]{Wang2}%
  \BibitemOpen
  \bibfield  {author} {\bibinfo {author} {\bibfnamefont {J.}~\bibnamefont {Wang}} \emph {et~al.},\ }\bibfield  {title} {\bibinfo {title} {T-square resistivity without umklapp scattering in dilute metallic bi$_2$o$_2$se},\ }\href@noop {} {\bibfield  {journal} {\bibinfo  {journal} {Nat. Commun.}\ }\textbf {\bibinfo {volume} {11}},\ \bibinfo {pages} {3846} (\bibinfo {year} {2020})}\BibitemShut {NoStop}%
\bibitem [{\citenamefont {Behnia}(2022)}]{Behnia}%
  \BibitemOpen
  \bibfield  {author} {\bibinfo {author} {\bibfnamefont {K.}~\bibnamefont {Behnia}},\ }\bibfield  {title} {\bibinfo {title} {On the origin and the amplitude of t-square resistivity in fermi liquids},\ }\href@noop {} {\bibfield  {journal} {\bibinfo  {journal} {Ann. Phys.}\ }\textbf {\bibinfo {volume} {534}},\ \bibinfo {pages} {2100588} (\bibinfo {year} {2022})}\BibitemShut {NoStop}%
\bibitem [{\citenamefont {Kumar}\ and\ \citenamefont {andf D.L.~Maslov}(2021)}]{Kumar}%
  \BibitemOpen
  \bibfield  {author} {\bibinfo {author} {\bibfnamefont {A.}~\bibnamefont {Kumar}}\ and\ \bibinfo {author} {\bibfnamefont {V.~Y.}\ \bibnamefont {andf D.L.~Maslov}},\ }\bibfield  {title} {\bibinfo {title} {Quasiparticle and nonquasiparticle transport in doped quantum paraelectrics},\ }\href@noop {} {\bibfield  {journal} {\bibinfo  {journal} {Phys. Rev. Lett.}\ }\textbf {\bibinfo {volume} {126}},\ \bibinfo {pages} {076601} (\bibinfo {year} {2021})}\BibitemShut {NoStop}%
\bibitem [{\citenamefont {Moll}\ \emph {et~al.}(2016)\citenamefont {Moll} \emph {et~al.}}]{Moll}%
  \BibitemOpen
  \bibfield  {author} {\bibinfo {author} {\bibfnamefont {P.}~\bibnamefont {Moll}} \emph {et~al.},\ }\bibfield  {title} {\bibinfo {title} {Evidence for hydrodynamic electron flow in pdcoo$_2$},\ }\href@noop {} {\bibfield  {journal} {\bibinfo  {journal} {Science}\ }\textbf {\bibinfo {volume} {351}},\ \bibinfo {pages} {1061–1064} (\bibinfo {year} {2016})}\BibitemShut {NoStop}%
\bibitem [{\citenamefont {Sebastian}\ and\ \citenamefont {Proust}(2015)}]{Sebastian}%
  \BibitemOpen
  \bibfield  {author} {\bibinfo {author} {\bibfnamefont {S.}~\bibnamefont {Sebastian}}\ and\ \bibinfo {author} {\bibfnamefont {C.}~\bibnamefont {Proust}},\ }\bibfield  {title} {\bibinfo {title} {Quantum oscillations in hole-doped cuprates},\ }\href@noop {} {\bibfield  {journal} {\bibinfo  {journal} {Ann. Rev. of Cond. Mat. Phys.}\ }\textbf {\bibinfo {volume} {6}},\ \bibinfo {pages} {411} (\bibinfo {year} {2015})}\BibitemShut {NoStop}%
\bibitem [{\citenamefont {Vool}\ \emph {et~al.}(2021)\citenamefont {Vool} \emph {et~al.}}]{Vool}%
  \BibitemOpen
  \bibfield  {author} {\bibinfo {author} {\bibfnamefont {U.}~\bibnamefont {Vool}} \emph {et~al.},\ }\bibfield  {title} {\bibinfo {title} {Imaging phonon-mediated hydrodynamic flow in wte$_2$},\ }\href@noop {} {\bibfield  {journal} {\bibinfo  {journal} {Nat. Phys.}\ }\textbf {\bibinfo {volume} {17}},\ \bibinfo {pages} {1216–1220} (\bibinfo {year} {2021})}\BibitemShut {NoStop}%
\bibitem [{\citenamefont {Gooth}\ \emph {et~al.}(2018)\citenamefont {Gooth} \emph {et~al.}}]{Jogi}%
  \BibitemOpen
  \bibfield  {author} {\bibinfo {author} {\bibfnamefont {J.}~\bibnamefont {Gooth}} \emph {et~al.},\ }\bibfield  {title} {\bibinfo {title} {Thermal and electrical signatures of a hydrodynamic electron fluid in tungsten diphosphide.},\ }\href@noop {} {\bibfield  {journal} {\bibinfo  {journal} {Nat. Commun.}\ }\textbf {\bibinfo {volume} {9}},\ \bibinfo {pages} {4093} (\bibinfo {year} {2018})}\BibitemShut {NoStop}%
\bibitem [{\citenamefont {Coulter}\ \emph {et~al.}(2018)\citenamefont {Coulter}, \citenamefont {Sundararaman},\ and\ \citenamefont {Narang.}}]{Pri}%
  \BibitemOpen
  \bibfield  {author} {\bibinfo {author} {\bibfnamefont {J.}~\bibnamefont {Coulter}}, \bibinfo {author} {\bibfnamefont {R.}~\bibnamefont {Sundararaman}},\ and\ \bibinfo {author} {\bibfnamefont {P.}~\bibnamefont {Narang.}},\ }\bibfield  {title} {\bibinfo {title} {Microscopic origins of hydrodynamic transport in the type-ii weyl semimetal wp$_2$},\ }\href@noop {} {\bibfield  {journal} {\bibinfo  {journal} {Phys. Rev. B}\ }\textbf {\bibinfo {volume} {98}},\ \bibinfo {pages} {115130} (\bibinfo {year} {2018})}\BibitemShut {NoStop}%
\bibitem [{\citenamefont {Huang}\ and\ \citenamefont {Lucas}(2021)}]{Lucas}%
  \BibitemOpen
  \bibfield  {author} {\bibinfo {author} {\bibfnamefont {X.}~\bibnamefont {Huang}}\ and\ \bibinfo {author} {\bibfnamefont {A.}~\bibnamefont {Lucas}},\ }\bibfield  {title} {\bibinfo {title} {Electron-phonon hydrodynamics},\ }\href@noop {} {\bibfield  {journal} {\bibinfo  {journal} {Phys. Rev. B}\ }\textbf {\bibinfo {volume} {103}},\ \bibinfo {pages} {155128} (\bibinfo {year} {2021})}\BibitemShut {NoStop}%
\bibitem [{\citenamefont {Liu}\ \emph {et~al.}(2023)\citenamefont {Liu} \emph {et~al.}}]{Liu}%
  \BibitemOpen
  \bibfield  {author} {\bibinfo {author} {\bibfnamefont {Y.}~\bibnamefont {Liu}} \emph {et~al.},\ }\bibfield  {title} {\bibinfo {title} {Gate-tunable multiband transport in zrte$_5$ thin devices},\ }\href@noop {} {\bibfield  {journal} {\bibinfo  {journal} {Nano Lett}\ }\textbf {\bibinfo {volume} {11}},\ \bibinfo {pages} {5334–5341} (\bibinfo {year} {2023})}\BibitemShut {NoStop}%
\end{thebibliography}%

\end{document}